\documentclass[preprints,article,accept,oneauthor,pdftex]{mdpi_mod} 

\Title{Cosmological Constant from Boundary Condition and Its Implications beyond the Standard Model}

\Author{Jan O. Stenflo $^{1,2}$}

\address{%
$^{1}$ \quad Institute for Particle Physics and Astrophysics, ETH  Zurich, CH-8093 Zurich, Switzerland; stenflo@astro.phys.ethz.ch\\
$^{2}$ \quad IRSOL Istituto Ricerche Solari ``Aldo e Cele Dacc\`o'', Universit\`a  della Svizzera Italiana, via Patocchi 57, CH-6605 Locarno, Switzerland}

\corres{Correspondence: stenflo@astro.phys.ethz.ch}

\newcommand{\aj}{Astron.~J.} 
\newcommand{\apj}{Astrophys.~J.} 
\newcommand{\apjl}{Astrophys.~J.~Lett.} 
 
\newcommand{\aap}{Astron.~Astrophys.} 
\newcommand{\araa}{Ann.~Rev.~Astron.~Astrophys.} 
\newcommand{\aapr}{Astron.~Astrophys.~Rev.} 
\newcommand{\mnras}{Monthly~Not.~Royal~Astron.~Soc.} 
\newcommand{\prd}{Phys.~Rev.~D} 
\newcommand{\prl}{Phys.~Rev.~Lett.} 
\newcommand{\jcap}{J.~Cosmology Astropart.~Phys.} 

\abstract{
Standard cosmology has long been plagued by a number of persistent problems. The origin of the apparent acceleration of the cosmic expansion remains enigmatic. The cosmological constant has been reintroduced as a free parameter with a value in energy density units that ``happens'' to be of the same order as the present matter energy density. There is an internal inconsistency with regards to the Hubble constant, the so-called $H_0$ tension. The derived value of $H_0$ depends on the type of data that is used. With supernovae as standard candles one gets a $H_0$ that is 4-5 $\sigma$ larger than the value that one gets from CMB (Cosmic Microwave Background) data for the early universe. Here we show that these problems are related and can be solved if the cosmological constant represents a covariant integration constant that arises from a spatial boundary condition, instead of being a new type of hypothetical physical field, ``dark energy'', as assumed by standard cosmology. The boundary condition only applies to the bounded 3D subspace that represents the observable universe, the hypersurface of the past light cone. 
}


\begin{document}

\section{Introduction}\label{sec:intro}

While the standard model of cosmology, which is often referred to as the concordance model or the $\Lambda$CDM model, has been highly successful in the modelling of a wealth of cosmological observations, it has not been able to provide resolutions to several fundamental problems. The cosmological constant $\Lambda$ had to be introduced as a free fitting parameter to allow modelling of the apparently accelerated cosmic expansion, which had been discovered unexpectedly from the observed redshift - brightness relation for standard candles in the form of supernovae type Ia \cite{stenflo-riessetal1998a,stenflo-perlmutteretal1999a}. Explanations have been searched
for in terms of some new field, referred to as ``dark energy'', a
source of repulsive gravity with an equation of state such as the
vacuum energy of quantum fields 
\cite{stenflo-Frieman_review2008,stenflo_durrer2011,stenflo-binetruy2013,stenflo-joyce_review2016,stenflo-amendola2018,stenflo-sabulsky2019}. 
A problem with this interpretation is that 
quantum field theory predicts a value of $\Lambda$ that is about 54
orders of magnitude larger than the observed value (not the often 
quoted 122 orders of magnitude, cf. \cite{stenflo-martin2012cc}). This is considered as one of the worst predictions in the history of physics. 
Although a variety of alternative explanations have been explored, e.g. 
\cite{stenflo-padmanabhan2017,stenflo-lombriser2019},
the cosmological constant has remained enigmatic. 

In the absence of any known constraint on the value of the
cosmological constant it has been argued 
that any value of $\Lambda$ may be physically permissible, 
that there may exist parallel universes where its value is different,
and that only universes where the value is very small are
conducive to the emergence of 
biological life \cite{stenflo-weinberg1987}. 
If however a unique constraint condition for $\Lambda$ could be identified, this 
``anthropic argument'' would become obsolete and the multiverse option
would be ruled out. 

Standard cosmology has persistently stuck to the interpretation that the cosmological constant is some kind of unspecified physical field that is yet to be discovered, in spite of unsuccessul decades in search for such a hypothetical field. Internal inconsistencies also plague standard cosmology, the most prominent of which --- generally referred to as the ``$H_0$ tension'' --- is related to the Hubble constant, a parameter that
establishes the fundamental scale for cosmological distances and for
the age of the universe. While the use of supernovae type Ia as
standard candles gives $H_{0,\,{\rm SN}} =73.2\pm
1.3$\,km\,s$^{-1}$\,Mpc$^{-1}$ \cite{stenflo-riessetal2018,stenflo-riess2021a}, the CMB
analysis gives  $H_{0,\,{\rm CMB}} =67.4\pm
0.5$\,km\,s$^{-1}$\,Mpc$^{-1}$ \cite{stenflo-planck2020a}. It is a disagreement that
  exceeds $4\sigma$. 

There is a similar tension in another fundamental distance scale, the
radius of the sound horizon $r_d$ at the ``drag epoch'', when the
baryons decouple from the photons. If we allow $r_d$ to be a free
fitting parameter, then BAO (Baryon Acoustic Oscillations)
observations for small redshifts give $r_d\approx 136\pm 3$\,Mpc, much
smaller than the CMB-derived value of $r_d =147.2\pm 0.3$\,Mpc
\cite{stenflo-arendse2020}. 

The tensions in $H_0$ and $r_d$ are however anti-correlated. There is
no significant tension in the $H_0\,r_d$ product 
\cite{stenflo-arendse2020}. Late-time modifications of the standard
model, e.g. by allowing the cosmological constant to vary, would only
affect $H_0$ but not 
$r_d$. Early-time modifications would require new physics to be
postulated beyond the standard model of particle physics, but this
would still fail to 
resolve the problem, because a reduction of $r_d$ would exacerbate a
tension in the $S_8$ parameter \cite{stenflo-jedamzik2021}. $S_8$ is
a measure of the matter clustering amplitude on a scale of $8/h$\,Mpc
(where $h$ is the Hubble constant in units of
100\,km\,s$^{-1}$\,Mpc$^{-1}$). 

The present paper aims to show that these enigmas and inconsistencies can be resolved if one abandons the standard dark-energy interpretation of the cosmological constant and instead treats it as a covariant integration constant that arises from a spatial boundary condition that only applies to the bounded hypersurface of the past light cone, the 3D subspace that represents the observable universe. With this interpretational framework, which keeps the formalism within the realm of general relativity (in contrast to many other more or less ad hoc alternative cosmologies that have been put forward), it is possible to resolve the outstanding enigmas and to predict the numerical values of the cosmological constant and the $H_0$ tension, while remaining consistent with CMB data and Big Bang nucleosynthesis. Without the use of adjustable parameters the predictions are found to agree with the observationally determined values within the observational uncertainties. 

Section 2 clarifies why the Einstein field equation for gravity and cosmology must contain a covariant integration constant. Section 3 discusses the fundamental distinction between the case when the scale factor $a$ is a function of proper time $t$ that can be measured by comoving clocks in 4D spacetime, and when it is a function of cosmological distance $r$ along light rays in a finite 3D subspace. Section 4 is devoted to the derivation of the expression and numerical value for the cosmological constant. Section 5 deals with two further implications beyond the standard model: the age of the universe, and a resolution of the $H_0$ tension. The conclusions are summarized in Section 6.

\section{Why the Metric Needs to be Constrained by a Boundary Condition}\label{sec:metricconstraint}

Einstein's equation for gravity is generally written with the expression for the spacetime metric on the left-hand side and the sources of matter and energy in the form of the stress-energy tensor $T_{\mu\nu}$ on the right-hand side. While mass-energy governs how spacetime should curve, the geometry determines how matter and radiation should move. 

The metric is constrained by the requirement of mass-energy conservation: the covariant divergence of the stress-energy tensor must be zero. This implies that the Bianchi identities for the metric must be satisfied. The solution of such differential constraints generally requires the introduction of integration constants that need to be fixed by boundary conditions. 

Integration of the Bianchi identities leads to the left-hand side of the Einstein field equation plus an integration constant that must be given in covariant form, a constant multiplied by the metric tensor $g_{\mu\nu}$, because the covariant divergence of such a term is zero. This is the form of the cosmological constant $\Lambda$ term, which Einstein \cite{stenflo-einstein17} introduced in 1917 for the mistaken purpose of obtaining a static cosmological solution. 

The $g_{\mu\nu}\Lambda$ term can be placed on either the left or the right-hand side of the equation, to be considered as a degree of freedom for either the metric field or for the stress-energy tensor. Standard cosmology interprets the $\Lambda$ term as a component of the stress-energy tensor $T_{\mu\nu}$ and refers to it as ``dark energy''. Instead of being an integration constant it is treated as a new kind of field, although there is no known physical basis for the existence of such a peculiar field (which needs to be repulsive with an equation of state $w\approx-1$). 

The choice between the two interpretations of the $\Lambda$ term, integration constant or physical field, is not just a matter of viewpoint but has profound consequences for both the conceptual and quantitative understanding and modelling of observational data. This makes it possible to  use observations to unambiguously  discriminate between the two possible choices.

\section{Scale Factor as a Function of Distance}\label{sec:findist}

\subsection{Measures of Time and Distance}\label{sec:measures}

Although space and time are tied together as 4D spacetime, which is the general arena where physical processes play out, there are fundamental differences between the temporal and spatial aspects in the cosmological context. Observers move in time but not in space, while light rays move in space but not in time. Cosmological observers are defined to be comoving, at rest with respect to  the spatial grid. The observable universe represents the static 3D subspace along light rays that can be seen by the observer at a given proper time (which defines the tip of the past light cone). 

The time scale that governs all dynamics is the proper time $t$ that can be measured by comoving clocks. It is the time scale of stellar evolution, structure formation, and the  density and temperature fluctuations of the CMB. 

All information that we receive from cosmological objects comes to us in the form of radiation, electromagnetic or gravitational. While the universe may be infinite in both space and time, the part of it that is accessible to observation represents a time-frozen 3D subspace of finite volume, the 3D hypersurface of the past light cone. The distance scale $r$ along the light cone is bounded by the Big Bang, where $r$ equals the radius $r_c$ of the causal or particle horizon. It is the distance where the scale factor $a$ goes to zero and the redshift $z$ goes to infinity. At the other end the distance scale is bounded where $r=0$ (location of the observer), because negative distances $r<0$ have no defined meaning.

\subsection{Interpretation of Redshift Observations}\label{sec:redshift}

A basic assumption of cosmological modelling is that the universe is spatially homogeneous and isotropic (on large scales), and that the expansion of space can be specified in terms of a scale factor $a(t)$, which only depends on the proper time $t$ that can be measured by comoving clocks (at rest with respect to the spatial grid). The FLRW (Fried\-mann-Lema\^itre-Robertson-Walker) models represent solutions of the Einstein equation in the form of scale factor $a$ vs. proper time $t$. The function $a(t)$ depends on the (spatially homogeneous) mass densities $\rho_M$ of matter, $\rho_R$ of radiation, and $\rho_\Lambda$ (the cosmological constant when expressed in mass density units). 

The function $a(t)$ is however not an observable. When we observe objects at different cosmological distances, like standard candles in the form of SN (supernovae) type Ia, we measure their redshifts $z$ and apparent brightnesses or received energy fluxes ${\cal F}$, to obtain the function $z$ vs. ${\cal F}$. To interpret this function in terms of the model parameters $\rho_{M,R,\Lambda}$ we need to relate the observational $z({\cal F})$ function to the theoretical $a(t)$ function. 

Redshift $z$ is directly related to scale factor $a$ via 
\begin{equation}\label{eq:azdef}
a=1/(1+z)\,.
\end{equation}
Because the energy flux ${\cal F}$ that is received from an object decreases with the square of the distance, it is related to the luminosity $L$ of the source through 
\begin{equation}
{\cal F}\equiv \frac{L}{4\,\pi\,d_L^{\,2}}\,,
\label{eq:appenflux}\end{equation}
which defines the so-called luminosity distance $d_L$. 
$L$ can be assumed to be known for standard candles through applications of the distance ladder with its intricate calibrations, based on parallaxes for relatively nearby objects and Cepheid variable stars for intermediate distances. Equation (\ref{eq:appenflux}) then gives us $d_L$ from the observed brightness ${\cal F}$. If the universe were not expanding, $d_L$ would equal the coordinate distance $r$. Because of the expansion, the relation between $d_L$ and $r$ depends on the scale factor $a$: 
\begin{equation}\label{eq:radl}
r=a(r)\, d_L\,.
\end{equation}

Note that in Equation (\ref{eq:radl}) the scale factor $a$ has been written as a function of distance $r$ instead of proper time $t$, because $t$ is not an observable, and it does not flow along $r$. While time dependence with dynamical evolution can be observed for astrophysical structures on small or intermediate scales, they are unobservable on the cosmic time scale that governs the evolution of the scale factor, because the human time scale is negligible in comparison with cosmic time. Therefore the observable universe in terms of the observed $z({\cal F})$ function is a time-frozen representation of the universe at the present epoch $t_0$ of human observers. 

Since the evolutionary cosmic time scale $t$ is not an observable, we need a mapping of the $a(t)$ evolution (which is unbounded  in the future direction) onto the bounded ($0< r<r_c$) distance scale, to obtain the $a(r)$ function that is part of Equation (\ref{eq:radl}). The problem is that the distance scale is not simply obtained by mathematical projection onto the light cone. Distances in space and time are only given a meaning in terms of a metric field, which represents a solution of the Einstein field equation for the given spacetime domain. The scale factor $a$ is a component of the metric tensor that is obtained by solving the field equation. In the finite and static 3D subspace that represents the observable universe the metric field can be subject to a boundary constraint that may be satisfied by a non-zero covariant integration constant, which would appear in the form of the $\Lambda$ term in the equation for $a(r)$. In contrast, because 4D spacetime is unbounded, no integration constant can get induced in the equation for $a(t)$.

\subsection{Meaning of the Hubble Constant}\label{sec:hubblemeaning}

The $a(t)$ function is needed to model cosmological evolution, in particular for numerical simulations of structure formation or for calculations of the evolution of the density and temperature fluctuations in the early universe. The most important observational signatures of this evolution are the acoustic peaks in the CMB (Cosmic Microwave Background) spectrum, which represent the anisotropies of the cosmic  radiation ``bath'', in which we are immersed. 

While the $a(t)$ function should be used for interpretations of the CMB spectrum, the $a(r)$ function is to be used for the redshift observations of supernovae. An inconsistency of standard cosmology, the so-called $H_0$ tension, arises when the different roles of the $a(t)$ and $a(r)$ functions are not accounted for (cf. Section \ref{sec:tension}). 

The $H_0$ tension has to do with the assumption of standard cosmology that the Hubble constant $H$ can be expressed as ${\dot a}/a$, where ${\dot a}\equiv {\rm d}a/{\rm d}t$ and $t$ represents proper time. When the $a(r)$ function is affected by a boundary-induced cosmological constant but the $a(t)$ function is not, the representation of the Hubble constant as ${\dot a}/a$ becomes invalid. 

Historically the Hubble constant was introduced to describe the observed linear relation between velocity $v$ of a galaxy that moves away from us and its distance $r$ by the equation $v=Hr$. The measured quantity was however redshift $z$ and not $v$. Since $v=cz$ for $v\ll c$ according to the Doppler effect, Hubble's law should be written $z=Hr/c$, or, in differential form, 
\begin{equation}\label{eq:hubdefdiff}
\frac{{\rm d}z}{{\rm d}r}=H/c\,.
\end{equation}
Since redshift $z$ is directly related to scale factor $a$ via Equation (\ref{eq:azdef}), one can define the Hubble constant in terms of the scale factor as 
\begin{equation}\label{eq:hubdefa}
H\equiv -\,\frac{c}{a^2} \,\frac{{\rm d}a}{{\rm d}r}\,\ne \,\frac{\dot a}{a}\,.
\end{equation}
The last expression, to the right of the $\ne$ sign, is the expression that standard cosmology uses for the Hubble constant. It is obtained from our definition in terms of ${\rm d}a/{\rm d}r$ if one does the substitution ${\rm d}r \to -c\,{\rm d}t/a$ (with the minus sign because $t$ decreases when $r$ increases). It is conceptually incorrect, because $\dot a$ represents a time derivative with respect to proper time and depends on the $a(t)$ function, while the relevant spatial derivative depends on the $a(r)$ function. The inequality happens when $\Lambda$ is different for the $a(t)$ and $a(r)$ functions.

\section{Derivation of the Value for $\Lambda$}\label{sec:valuelambda}

Standard FLRW cosmology is based on the assumption that there is spatial homogeneity of the universe on large scales, not only for the matter and radiation densities $\rho_M$ and $\rho_R$, but also for the scale factor $a$. All spatial gradients are prescribed to be zero, there is only evolution. The large-scale metric is assumed to be diagonal and isotropic. In the following we will add the assumption of spatial flatness, which makes the expressions simpler and more transparent. As observations show that the spatial curvature is zero within the observational uncertainties \cite{stenflo-planck2020a}, there is no need to complicate the expressions by including the curvature term. 

With these assumptions $g_{tt}=-1$ and $g_{xx}= g_{yy}= g_{zz}=a^2(t)$ in terms of Cartesian coordinates that are orthogonal to the time axis. The scale factor is thus prescribed to be independent of the spatial coordinates and depends exclusively on proper time $t$. The same metric can also be expressed in terms of spherical coordinates $r$, $\theta$, and $\phi$, as is common in FLRW cosmological modeling. Then, with spatial flatness, $g_{rr}=a(t)^2$, $g_{\theta \theta}=a(t)^2\,r^2$, and $g_{\phi \phi}=a(t)^2\,r^2\sin^2\theta$.

\subsection{Spatial Perturbations of the FLRW Metric}\label{sec:flrwperturb}

In the FLRW formulation the 4D problem is by assumption reduced to a 1D problem: The large-scale evolution is fully described by a scale factor $a(t)$ that only varies with the proper time $t$ that can be recorded by local, comoving clocks. Because there is no dependence on distance in the direction orthogonal to the time axis, the $a(t)$ function can be readily obtained as a 1D solution of the Einstein equation while accounting for energy conservation for matter and radiation. 

Our aim now is to determine the scale factor as a function of distance $r$ in the observable universe (along the past light cone). The $a(r)$ function relates the scale factor between the positions of non-local objects, which are separated from the comoving observer by different distances within a bounded $r$ domain. The FLRW prescription for a metric of 4D spacetime that only depends on local, unbounded proper time is too restrictive and limited to be of use to account for the bounded distance scale. It becomes necessary to relax the FLRW requirement of zero spatial gradients by allowing for small-amplitude vacuum perturbations of the metric in the form of a ``probe'' field $h(r)$.  

The metric of our minimally extended, perturbed version of FLRW has the following form (in units where $c=1$): 
\begin{equation}
{\rm d}s^2 = - {\rm d}t^2 +a(t)^2 \,[1+h(r)]\,[\,{\rm d}r^2+ \,r^2\,(\,{\rm  d}\theta^2 + \,\sin^2\theta\,\,{\rm
  d}\phi^2\,)\,]\,,
\label{eq:flrw}\end{equation}
where $h(r)\ll 1$ represents the small-amplitude probe field. It is needed to define consistent boundary conditions, which relate to the static 3D subspace that constitutes the observable universe. $h$ is by design isotropic with respect to the observer. The size of the 3D subspace in which the probe field is confined determines the scale of the fluctuations. The $a(t)$ function, which governs the evolution of the CMB spectrum and is not constrained by any boundaries, is not affected.

\subsection{Solution of the Einstein Equation in the Presence of  Spatial Perturbations}\label{sec:soleinspatial}

It is convenient to express the Einstein equation in the form 
\begin{equation}
R_{\mu\nu} -\,g_{\mu\nu}\Lambda =8\pi \,G\,S_{\mu\nu}
\label{eq:einst}\end{equation}
in units where $c=1$. 
\begin{equation}
S_{\mu\nu} =T_{\mu\nu} -{\textstyle{\frac{1}{2}}} \,g_{\mu\nu} T 
\label{eq:smunumr}\end{equation}
contains the energy-momentum sources. The metric is represented by the Ricci tensor $R_{\mu\nu}$ and the cosmological $\Lambda$ term. For the signature of the metric we use the convention ($-++\,+$). 

For our purpose of exploring the relation between $\Lambda$ and the spatial boundary conditions it is sufficient to focus our attention on the $R_{\theta\theta}$ component of the Ricci tensor, because the expression for the other components  is analogous. To first order in the perturbation $h$ one finds 
\begin{equation}
R_{\theta\theta} -\,g_{\theta\theta}\,\Lambda \approx\,g_{\theta\theta}\,\biggl[\,\frac{\ddot a}{a}\,+\,2\left(\frac{\dot a}{a}\right)^{\!\!2}\,\biggr]-\,\textstyle{\frac{1}{2}}\,r^2\,\,\nabla^2 \,h\,-\,g_{\theta\theta}\,\Lambda\,=\, 4\pi\,G(\rho -p) \,g_{\theta\theta}\,.
\label{eq:asolut1}\end{equation}
The expression for $S_{\theta\theta}={\textstyle\frac{1}{2}}(\rho-p) \,g_{\theta\theta}$ has been used, where $\rho$ is the mass density and $p$ the pressure. 

Because the spatial and temporal coordinates in this 4D spacetime representation are orthogonal to each other, the $t$-dependent terms assume their values independent of the spatial $\nabla^2 \,h$ term. There is no boundary-induced integration constant that could be the source of $\Lambda$, because the $t$ and $r$ coordinates in infinite 4D spacetime are unbounded. If one refrains from postulating the existence of some ``dark energy'' field, then $\Lambda$ in Equation (\ref{eq:asolut1}) is zero. The time-dependent terms decouple to form an equation that is independent of the spatial perturbation: 
\begin{equation}
\frac{\ddot a}{a}\,+\,2\left(\frac{\dot a}{a}\right)^{\!\!2}=\,\frac{1}{a^3}\,\frac{{\rm d}^2 a}{{\rm d}\eta^2}\,+\,\frac{1}{a^4}\left(\frac{{\rm d}a}{{\rm d}\eta}\right)^{\!\!2}= \,4\pi\,G(\rho -p)\,.
\label{eq:tetarho}\end{equation}
It represents the evolutionary equation for the unperturbed background, a standard Friedmann universe without a cosmological constant. The time dependence of the scale factor has been expressed in two equivalent forms, one with respect to proper time $t$, the other with respect to conformal time $\eta$, which is defined by ${\rm d}\eta \equiv{\rm d}t/a$. 

Let us now rewrite Equation (\ref{eq:asolut1}) by rearranging the terms and expanding the expression for $g_{\theta\theta}$: 
\begin{equation}
(1+h)\Bigl[\,\frac{1}{a^3}\,\frac{{\rm d}^2 a}{{\rm d}\eta^2}\,+\,\frac{1}{a^4}\left(\frac{{\rm d}a}{{\rm d}\eta}\right)^{\!\!2}\!- \,4\pi\,G(\rho -p) \,\Bigr]\,= \,\frac{1}{2a(t)^2}\,(\,\nabla^2\, h\,+\,2a(t)^2 \Lambda\, h \,)\,+\,\Lambda\,.
\label{eq:asolut2}\end{equation}
The left-hand side vanishes because of Equation (\ref{eq:tetarho}). While this implies that the right-hand side must also vanish, it does not lead to any constraint on the magnitude of  $\Lambda$. Since the 4D representation is unbounded, it does not offer the possibility of useful boundary conditions. 

The situation changes when we impose the restriction to the light cone and thereby reduce the domain from infinite 4D spacetime to the bounded 3D subspace that represents the observable universe.

\subsection{Restriction to the Light Cone}\label{sec:restlcone}

The observer is by definition located at the spacetime tip of the past light cone. The observable universe represents the static and bounded 3D hypersurface of this null cone. It is bounded between $r=0$ (where the redshift $z=0$) and $r=r_c$ (where the redshift $z$ becomes infinite). The distance $r_c$ to the causal horizon increases with time, which makes $r_c$ a function of scale factor $a$. For any given value of $a$, the coordinate distance $r$ along a hypersurface of constant $a$ has a limiting value $r_c(a)$ that defines the causally connected domain. Any disturbance that originates in the Big Bang can spread to a maximal distance $r_c(a)$ by the time $t$ when the scale factor has reached the value $a(t)$. 

On the light cone all spacetime intervals ${\rm d}s$ vanish (which can be seen as an expression of the property that proper time does not flow along light rays). This implies a differential relation between the $r$ and $\eta$ coordinates. According to Equation (\ref{eq:flrw}) the ${\rm d}s^2=0$ restriction  implies that 
\begin{equation}
{\rm d}\eta^2 =(1+h)\,{\rm d}r^2
\label{eq:ds0restrict}\end{equation}
in the radial direction. Implementing this in Equation (\ref{eq:asolut2}) gives 
\begin{equation}
\frac{1}{a^3}\,\frac{{\rm d}^2 a}{{\rm d}r^2}\,+\,\frac{1}{a^4}\left(\frac{{\rm d}a}{{\rm d}r}\right)^{\!\!2}\!- \,4\pi\,G(\rho -p) \,-\Lambda= \frac{1}{2a(t)^2}\Bigl[\nabla^2 \,h(r)\,+\,2a(t) \,\Lambda\, h(r) \Bigr]\,.
\label{eq:asolut3}\end{equation}
We have not included the term $- \,4\pi\,G(\rho -p)\,h(r)$, because $h$ is treated as a probe field that represents vacuum fluctuations of the metric and therefore does not couple to the material sources. 

Note that because ${\rm d}s^2=0$ is a differential constraint, it only transforms the differential expressions on the left-hand side, but not the proper time $t$ in $a(t)$ on the right-hand side. Since proper time and distance are independent coordinates, the zero points of their scales are unrelated to each other. Only when we introduce an observer (by definition at $r=0$), who exists when the age of the universe is $t=t_0$ and the scale factor is $a(t_0)$, the equation becomes physically well defined. Distances $r$ only have a meaning relative to an observer. 

Proper time does not vary with $r$ in contrast to the formal ``look-back time'', which does not represent a dynamically relevant time scale. Therefore the scale factor $a$ is allowed to vary with distance $r$ on the left-hand side of the equation, while $a(t)$ on the right-hand side remains constant because $t$ is kept fixed to represent the proper time of the tip of the light cone, which defines the $r$ scale that is used for the left-hand side.  

In the limit of vanishing spatial perturbations $h$ the right-hand side of Equation (\ref{eq:asolut3}) is zero, and we are left with an equation for the unperturbed backgound. It shows how the scale factor $a$ varies with distance $r$ from the perspective of the observer, who is located at $r=0$ at proper time $t_0$. From this vantage point in spacetime the observable universe represents a time-frozen ``snapshot'', a bounded 3D subspace that is embedded in the unbounded 4D universe. 

The right-hand side of Equation (\ref{eq:asolut3}) serves both to define the light cone to which the left-hand side refers and to constrain the value of $\Lambda$. The proper time $t$ there defines the temporal location for the tip of the light cone (while the spatial location of the tip and the observer is arbitrary). For time $t=t_0$ (the present age of the universe, when human observers exist), the scale factor has the value $a_0=a(t_0)$. The standard convention in cosmological modelling is to normalize the scale factors so that $a(t_0)\equiv 1$. 

For this particular value of $a$, which represents our time in cosmic history, the perturbation $h$ is confined within the bounded distance interval $0\le r \le r_c(a_0)$. Equation (\ref{eq:asolut3}) is only satisfied if the right-hand side of the equation with $a=a_0$ is zero, i.e., when 
\begin{equation}
\nabla^2 \,h(r)\,+\,2\Lambda\, h(r) =0\,.
\label{eq:hconstraint}\end{equation}
This equation is subject to a boundary condition, which constrains the value of $\Lambda$.  Equation (\ref{eq:asolut1}) did not admit such a  possibility, because it represents unbounded 4D spacetime. It is the light cone restriction that generates a bounded 3D subspace in which a non-zero $\Lambda$ can get induced. This is the reason why Equation (\ref{eq:asolut3}) for the $a(r)$ function can contain a non-zero cosmological constant, while Equation (\ref{eq:tetarho}) for the $a(t)$ function cannot. 

Equation (\ref{eq:hconstraint}) is a 3D equation for a spherical coordinate system, but since there is only dependence on the $r$ coordinate because of isotropy, it is more convenient to express it in a 1D form that eliminates the coordinate singularity at $r=0$ and makes the perturbation symmetric with respect to the two boundaries (0 and $r_c$). This is accomplished if we replace $h$ with $\psi$, where 
\begin{equation}
\psi\equiv r\,h\,.
\label{eq:psirh}\end{equation}
Inserting this in Equation (\ref{eq:hconstraint}) we get  
\begin{equation}
\frac{{\rm d}^2\psi}{{\rm d}r^2} +\,2\Lambda\,\psi\,=0\,.
\label{eq:1dhelmh}\end{equation}

\subsection{Implementation of the Spatial Boundary Condition to Obtain the Value for $\Lambda$}\label{sec:implementbc}

Equation (\ref{eq:1dhelmh}) is satisfied for Fourier modes with wave number $k_\Lambda$ and wavelength $r_\Lambda$, which are related to the cosmological constant through 
\begin{equation}
\sqrt{2\Lambda}=k_\Lambda\equiv\frac{2\pi}{r_\Lambda}\,.
\label{eq:klambda}\end{equation}
The wavelength $r_\Lambda$ defines a length scale, which can be expected to be related to the only  available cosmological length scale, the size $r_c$ of the bounded 3D subspace that represents the observable universe. 
We may therefore expect $r_\Lambda$ to be proportional to $r_c$ with a value for the proportionality constant that gets fixed by a boundary condition. 

At present the choice of boundary condition needs to be based on rather general symmetry arguments, because it does not follow from any yet identified equation. Regardless of its plausibility, a given choice will only be acceptable if it can be validated by observational data. An obvious choice is that the perturbation $\psi$ vanishes at both boundaries. It vanishes by definition at the location of the observer ($r=0$) because of Equation (\ref{eq:psirh}). A symmetric condition would then require that $\psi$ vanishes at the other boundary as well, at $r=r_c$, the beginning of time. 

The modes with the longest wavelengths (and lowest energies) that satisfy this condition are the ones for which the size $r_c$ of the bounded $r$ domain equals either half or a single wavelength. As shown in the next subsection, only the single-wavelength mode satisfies the observational constraints, but it does so with high precision (within 2\,\%\ or $2\sigma$).  This implies that 
\begin{equation}
r_\Lambda =r_c\,.
\label{eq:rlamrc}\end{equation}
In contrast the half-wavelength choice would give a value for $\Lambda$ that is four times smaller than the observed value. Other half-integer multiples would similarly be rejected by the observations by a huge margin. The observational data unambiguosly direct us to the single-wavelength choice of Equation (\ref{eq:rlamrc}). 

When the boundary condition that is defined by  
Equation (\ref{eq:rlamrc}) is inserted in Equation (\ref{eq:klambda}),  it 
gives a definite, predicted value for the cosmological constant: 
\begin{equation}
\Lambda =\Lambda_0/r_c^2\,,
\label{eq:lamrc}\end{equation}
where 
\begin{equation}
\Lambda_0 =2\,\pi^2\,.
\label{eq:lamrho}\end{equation}

The value of $\Lambda$ is constant throughout the observable universe, for the entire domain $0<r<r_c$, and therefore corresponds to an equation of state $w=-1$ if interpreted as an energy density. $\Lambda$ varies with epoch of the observer (but not with look-back time), because $r_c$ is a function of proper time $t$. This variation is however unobservable (because human time scales are insignificant compared to the age of the universe). If we use dimensionless spatial coordinates by expressing $r$ in units of $r_c$, then $\Lambda$ becomes equal to $\Lambda_0$ and remains constant, not only for all of observable space but also throughout all of cosmic history.

\subsection{Observational Validation}\label{sec:validation}

Equation (\ref{eq:lamrc}) represents a definite theoretical prediction for the magnitude of the cosmological constant $\Lambda$, which needs to be validated through comparison with the observationally determined value. To make this comparison explicit we first need to convert $\Lambda$ to the more convenient dimensionless form $\Omega_\Lambda$ that is used for model fits of cosmological data, and provide an explicit expression for the limiting distance scale $r_c$ (the radius of the particle horizon). 

Mathematically, the solution $a(r)$ for the $r$ dependence of the scale factor is obtained in the same way as the solution for $a(t)$, and is  
\begin{equation}\label{eq:avsrsol}
H(r)^2 = \left(\frac{8\pi\, G}{3\,c^2}\right)
[\,\rho_M/a(r)^3 +\rho_R/a(r)^4 \,]+\,\frac{c^2\Lambda}{3}\,. 
\end{equation}
The difference is that the Hubble constant is not represented by ${\dot a}/a$, but by $c\,{\rm d}z/{\rm d}r$ as in Equation (\ref{eq:hubdefdiff}) or in terms of $a$ as in Equation (\ref{eq:hubdefa}). Also, the $\Lambda$ term only appears in the expression that governs $a(r)$, not in the expression for $a(t)$, because the proper time scale is not bounded, in contrast to the $r$ scale in Equation (\ref{eq:avsrsol}). 

The next step is to introduce dimensionless parameters $\Omega_{M,R,\Lambda}$ through the definitions 
\begin{equation}\label{eq:omrho}
\Omega_{M,R} \equiv\frac{8\pi\,G}{3\,c^2
    H_0^2}\,\,\rho_{M,R}
\end{equation}
and 
\begin{equation}\label{eq:omlamdef}
\Omega_\Lambda\equiv\frac{c^2\Lambda}{3\,H_0^2}\,.
\end{equation}
This gives us 
\begin{equation}\label{eq:hh0solut}
H =H_0\,E(z)\,,
\end{equation}
where 
\begin{equation}\label{eq:h2solut}
E(z)\equiv \,\sqrt{\,\Omega_M(1+z)^3\,+\,\Omega_R(1+z)^4\,+\,\Omega_\Lambda\,}\,.
\end{equation}
Equations (\ref{eq:hh0solut}) and (\ref{eq:h2solut}) imply the normalization of the $\Omega_{M,R,\Lambda}$ coefficients 
\begin{equation}\label{eq:omnorm}
\Omega_M+\Omega_R+\Omega_\Lambda=1\,.
\end{equation}

The definition of the Hubble constant as $c\,{\rm d}z/{\rm d}r$ (cf. Equation (\ref{eq:hubdefdiff})) then allows us to calculate the distance $r$ to an object at a given redshift $z$: 
\begin{equation}\label{eq:rzrel}
r(z) = \int\! {\rm d}r = \frac{c}{H_0}\int_0^z\frac{{\rm d}z^\prime}{E(z^\prime)}\,\,.
\end{equation}
The limiting distance $r_c$, the causal horizon radius, is where the redshift $z$ becomes infinite: 
\begin{equation}\label{eq:rinfty}
r_c =r(\infty)\,.
\end{equation}

It is convenient to express $r_c$ in dimensionless form $x_c$, in units of the Hubble radius $c/H_0$: 
\begin{equation}\label{eq:xcrch0c}
x_c\equiv r_c\,H_0/c \,.
\end{equation}
From Equation (\ref{eq:rzrel}) we see that 
\begin{equation}\label{eq:xcexpress}
x_c = \int_0^\infty\frac{{\rm d}z}{E(z)}\,\,.
\end{equation}

Inserting the expression for $\Lambda$ from Equation (\ref{eq:lamrc}) into  the defining Equation (\ref{eq:omlamdef}) for $\Omega_\Lambda$ and using the definition for $x_c$ from Equation (\ref{eq:xcrch0c}), we get 
\begin{equation}\label{eq:defomlam2}
\Omega_\Lambda = \frac{2}{3}\,\left(\frac{\pi}{x_c}\right)^{\!\!2}\,. 
\end{equation}
The identical expression has been obtained previously via more heuristic derivations \cite{stenflo-s2018ccb,stenflo-s2020jpco}, which are nevertheless based on the same kind of boundary condition that is being used as the source of an integration constant in the form of the $\Lambda$ term. 

Since $x_c$ depends on $\Omega_\Lambda$ according to Equation (\ref{eq:xcexpress}), the solution cannot be expressed in closed algebraic form but needs to be obtained numerically. The apparent dependence on the density parameters $\Omega_M$ and $\Omega_R$ is not a real dependence, because the value of $\Omega_R$ is fixed by the observed temperature of the CMB radiation. The value of $\Omega_M$ then follows from the spatial flatness condition of Equation (\ref{eq:omnorm}) for a given choice of $\Omega_\Lambda$: 
\begin{equation}\label{eq:ommrel}
\Omega_M =1-\Omega_R-\Omega_\Lambda\,.
\end{equation}
These relations imply that the $E(z)$ function in Equation (\ref{eq:xcexpress}) only depends on $\Omega_\Lambda$ but not on $\Omega_{M,R}$. 

Equation (\ref{eq:defomlam2}) therefore defines a unique value for $\Lambda$, a predicted value that is obtained without any adjustable parameters. The solution that follows from our boundary condition is $x_c\approx 3.13$ and $\Omega_\Lambda\approx 0.671$. This is to be compared with the observationally based values of $\Omega_\Lambda =0.685\pm 0.007$ 
according to Planck Collaboration et al. \cite{stenflo-planck2020a} and $0.669\pm 0.038$ from
analysis of supernovae data by the Dark Energy Survey project
\cite{stenflo-DESAbbott2019a}. We note that the theoretical prediction
agrees with the Planck value within 
2\,\%\ or $2\sigma$ and is in even closer agreement with the DES value.

\section{Implications beyond the Standard Model}\label{sec:implsc}

The nearly perfect agreement between the observed and predicted values for the cosmological constant can be seen as a validation of the theory and a resolution of the long-standing problem of the origin of the $\Lambda$ term. At the same time the cosmological coincidence problem is made irrelevant. Our epoch in cosmic history is not special because of the particular value that the cosmological constant is observed to have. There is no violation of the Copernican principle. 

The theory has further implications beyond the standard model. Here we highlight two of them: the age of the universe, and a resolution of the $H_0$ tension.

\subsection{Age of the Universe}\label{sec:age}

It is commonly believed that the age of the Universe that is given by standard cosmology, 13.8\,Gyr, is known with high accuracy and is not in doubt.  However, this bypasses the fundamental fact that this age value is model dependent, and that the few direct measurements that are currently available actually slightly favor a larger value.

The age of the universe is readily obtained from the solution of the scale factor $a$ as a function of proper time $t$ that can be measured by a comoving clock. If one uses  the notations and concepts of standard cosmology, then the equation that governs this $a(t)$ function looks the same as Equation (\ref{eq:avsrsol}) that governs $a(r)$. In our case, however, the cosmological constant is only relevant for the $a(r)$ function that represents our bounded and static observable 3D subspace, not for $a(t)$, which represents the evolution of a comoving region in the 4D universe. Furthermore, it is conceptually incorrect to use notation $H$ to represent ${\dot a}/a$, because the Hubble constant depends on $a(r)$ for our observable universe but not on proper time $t$, as explained in Section \ref{sec:hubblemeaning}.  

With these conceptual differences the equation that governs $a(t)$ can be expressed as 
\begin{equation}\label{eq:atsolut}
{\dot a(t)}^2 = a(t)^2\left(\frac{8\pi\, G}{3\,c^2}\right)
[\,\rho_M/a(t)^3 +\rho_R/a(t)^4 \,]\,. 
\end{equation}
The density parameters $\rho_{M,R}$ have the same meanings and identical magnitudes as the corresponding parameters in Equation (\ref{eq:avsrsol}) for the $a(r)$ function. They represent the mass densities of matter and radiation at the present epoch $t_0$. As before, the scale factor is  assumed to be normalized to the value that it has at present: $a(t_0)\equiv 1$. 

The present age of the universe, $t_0$, is then obtained through straightforward integration of Equation (\ref{eq:atsolut}): 
\begin{equation}\label{eq:age1}
t_0=\int_0^{t_0} \!{\rm d}t =\left(\frac{3\,c^2}{8\pi\, G}\right)^{\!\!1/2} \!\int_0^1 \frac{{\rm d}a}{a \,(\rho_M/a^3 +\rho_R/a^4 ) ^{1/2}}\,.
\end{equation}

The solution of this integral expression can be obtained more conveniently after the $\rho$ parameters have been converted into the dimensionless $\Omega$ forms via the defining Equation (\ref{eq:omrho}). The result is 
\begin{equation}\label{eq:age2}
  t_0=\frac{1}{H_0} \int_0^1 \frac{{\rm d}a}{a \,(\Omega_M/a^3 +\Omega_R/a^4 ) ^{1/2}}\,\approx \frac{1}{H_0\sqrt{1-\Omega_\Lambda}} \int_0^1 \sqrt{a}\,\,{\rm d}a \, =\frac{2}{3H_0\sqrt{1-\Omega_\Lambda}}\,.
\end{equation}
Note that the present Hubble constant $H_0$ appears in this expression, but not because the Hubble constant has any conceptual relation to $t_0$. It is exclusively because $H_0$ is used in Equation (\ref{eq:omrho}) as a mathematical parameter that defines the conversion relation between the $\rho_{M,R}$ and $\Omega_{M,R}$ parameters. 

The second, approximate equality in Equation (\ref{eq:age2}) has been obtained by disregarding the contribution $\Omega_R$ due to radiation, which is $\ll \Omega_M$. The factor $\sqrt{1-\Omega_\Lambda}$ then appears as a consequence of Equation (\ref{eq:ommrel}). The approximation allows us to obtain a simple algebraic expression for the age $t_0$ in terms of the Hubble time $1/H_0$ and the cosmological constant. Without a cosmological constant one recovers the standard result that the age is 2/3 of the Hubble time when the universe is matter dominated. The presence of the $\Lambda$ term changes this result by the factor $1/\sqrt{1-\Omega_\Lambda}$. 

In our actual numerical evalution we have not made use of this approximation but retained the $\Omega_R$ term with its observed magnitude (obtained from the measured CMB temperature). However, the effect of $\Omega_R$ on the value of $t_0$ turns out to be smaller than one per mille and is thus insignificant. 

The numerical solution gives $t_0 =15.52$\,Gyr. The theoretical value $\Omega_\Lambda=0.671$ that comes from the solution of Equation (\ref{eq:defomlam2}) has been used together with the supernovae value for $H_0$, 73.2\,km\,s$^{-1}$\,Mpc$^{-1}$ \cite{stenflo-riessetal2018,stenflo-riess2021a},
because it has been derived in a way that is consistent with the present cosmological framework (within which $\Lambda$ is an integration constant), in contrast to the value that has been
obtained through CMB parameter fitting with the standard model. This is clarified in the next subsection that deals with the resolution of the $H_0$ tension.

The new age value is to be compared with the age 13.80\,Gyr that has been derived by Planck Collaboration et al. \cite{stenflo-planck2020a}. It is based on using the Planck values $\Omega_\Lambda =0.685$ and $H_0 =67.4\pm 0.5$\,km\,s$^{-1}$\,Mpc$^{-1}$ together with the dark-energy assumption of standard cosmology that the $a(t)$ and $a(r)$ functions are governed by the same $\Omega_\Lambda$ term. Including the $\Omega_\Lambda$ term in the solution for $a(t)$ adds an exponential contribution to the cosmic expansion that takes effect when one comes close to the present epoch. This accelerated expansion shortens the time scale and leads to a (formal) age value that is 1.7\,Gyr shorter than the proper, dynamically relevant age. 

The larger age relieves some existing tension between the age of the universe and the ages of the oldest stars. For a careful age comparison one needs to account for the time that it took to form the first stars after the Big Bang. First generation stars, so-called pop.~III stars with essentially zero
metallicity, have not yet been identified. The oldest, low-metallicity
stars that have been observed are second-generation pop.~II stars
in the halo of our galaxy. Numerical simulations indicate that the
rate of GC (globular cluster) formation may have peaked about
0.4--0.6\,Gyr after the Big 
Bang \cite{stenflo-trenti_etal2015,stenflo-naoz2006}. 

A reasonable estimate for the upper stellar age limit is therefore $t_0 -0.4$\,Gyr instead of  $t_0$ by itself. It is 13.4\,Gyr in the case of standard cosmology. The existence of a single star with a significantly larger reliable age would prove standard cosmology to be inconsistent. 

The most prominent stellar case that seems to violate the age limit of standard cosmology is represented by the so-called ``Methuselah'' star HD
140283. It is classified as a pop.~II halo subgiant at a distance of
202\,ly with a metallicity that is 250 times less than that of the
Sun. Stellar evolution modeling with
isochrones gives the age $14.46\pm 0.8$\,Gyr \cite{stenflo-bond2013a}. It violates the mentioned age limit by 1.1\,Gyr, but in terms of the error bars this is little more than one
$\sigma$. 

It has been claimed that the ages of globular clusters (GC) support the conventional age value for the universe. A GC age of $13.32\pm 
0.5$\,Gyr has been reported \cite{stenflo-valcin_etal2020}, which represents an average over  38 clusters that have
metallicity [Fe/H]$<-1.5$, but it was obtained after applying an ad hoc prior that
excludes values larger than 15\,Gyr \cite{stenflo-valcin_etal2020}. Because there is an age  distribution 
within the used subset, which exceeds what may be expected from the quoted
measurement uncertainties, the ad hoc exclusion of values above 15\,Gyr artificially truncates the intrinsic distribution. The average value that is extracted from such a distribution is therefore not representative of
the upper age limit of the clusters.  This gives reason to expect that the upper GC age limit lies significantly higher than the quoted value. Depending on the school of thought with respect to GC
formation \cite{stenflo-forbes_etal2018,stenflo-krumholz_etal2019} 
the GC age limit may indeed be consistent with a significantly larger
age of the universe.

These examples demonstrate that the current uncertainties in stellar age values are not yet small enough to conclusively discriminate between the two cosmological frameworks. The error bars are however expected to come down in the future. 

A new and independent avenue for the
determination of the ages of individual stars is that of
asteroseismology. The
detections by the CoRoT and Kepler satellite missions of oscillating
modes for many thousands of stars across the HRD have provided 
a rich data base for surveys of the age structure of the Milky Way
\cite{stenflo-silvaaguirre2016}. It has been used to
determine the vertical age gradient for the red giant stars in the
Galactic disk \cite{stenflo-casagrande2016a}. While the
majority of stars in the sample have ages in the range 1--6\,Gyr,
the tail of the age distribution extends well beyond 14\,Gyr, although
the sample is believed to be representative of the Galactic disk and
not of the halo. We do not yet know whether the extended tail is
exclusively an artefact of the uncertainties. 

Fortunately there is another testing ground, the enigmatic $H_0$ tension, which already now allows an unambiguous discrimination between the two cosmological frameworks. The $H_0$ tension represents an inconsistency of standard cosmology, which disappears within our alternative  framework, as explained in the next subsection.

\subsection{$H_0$ Tension}\label{sec:tension}

While the tension in the Hubble constant has received much attention in recent years, it has been pointed out \cite{stenflo-arendse2020}
that the $H_0$ tension should 
not be considered in isolation, because there is an anti-correlated
tension in the parameter $r_d$. This parameter represents the radius
of the sound horizon 
at the epoch when the baryons decouple from the photon drag force and
become free to cluster and form galactic structures. There is an imprint of
the $r_d$ scale on the observed galaxy distribution because of 
the Baryon Acoustic Oscillations (BAO). As $r_d\approx 1.02\, r_\star$
it is closely linked to the radius $r_\star$ of the sound horizon at
the epoch of hydrogen recombination, which governs the angular scale
of the acoustic peaks in the CMB spectrum. 

The value of $r_\star$ (as well as that of $r_d$) depends only on the
physics of the early universe, which is almost independent of the
cosmological model, because all effects of $\Lambda$ and spatial
curvature were insignificant before the $r_d$ and $r_\star$ 
epochs. The value of $r_\star$ is therefore predominanty constrained
by the standard model of 
particle physics and is found to be $r_\star \approx144.4\pm
0.3$\,Mpc, while the closely related scale
$r_d \approx 147.2\pm 0.3$\,Mpc \cite{stenflo-planck2020a}. With
these values the Planck CMB modelling results in a Hubble
constant $H_0 =67.4\pm 0.5$\,km\,s$^{-1}$\,Mpc$^{-1}$, significantly
smaller than $H_0 =73.2\pm 1.3$\,km\,s$^{-1}$\,Mpc$^{-1}$ that is obtained
from direct distance measurements in the nearby universe
\cite{stenflo-riessetal2018,stenflo-riess2021a}. It is this 
discrepancy between $H_0$ as determined from the
near and distant universe that is referred to as the $H_0$ tension. 

There is however no significant tension for the product $H_0\,r_d$ \cite{stenflo-arendse2020}. If one allows
$r_d$ to be a free parameter rather than being fixed by known physics,
the product $H_0\,r_d$ is constrained by the observed angular scale in
the CMB or the BAO, but the individual factors $H_0$ and $r_d$ are not. This
degeneracy can be broken by combining BAO with supernovae (SN)
observations. The result is a low value for $r_d$, typically
$r_d\approx 136\pm 3$\,Mpc (depending somewhat on the choice of
joint data sets), together with the high supernovae
value for $H_0$. The same kind of conclusion would be obtained from
analysis of CMB data, if one would allow $r_\star$ to be treated as a
free parameter. 

Attempts to find the origin for such a low value of $r_d$
in terms of new physics have been unsuccessful. It has been pointed out \cite{stenflo-arendse2020} that late-time 
modifications (like dark energy with a varying equation of state) 
only affects $H_0$ but not $r_d$ and therefore only makes things worse. All
proposed early-time modifications (before the CMB epoch) through the
introduction of new kinds of fields or modifications of general relativity
can be shown to be insufficient for a resolution of the
tension \cite{stenflo-arendse2020,stenflo-jedamzik2021}. The $H_0$ tension may be
slightly reduced by raising the value of the $\Omega_M\,
h^2$ parameter, but the potential of this change is limited, because
it would exacerbate the tension in the $S_8$ parameter that has
been revealed by the weak lensing surveys
\cite{stenflo-abbott2018,stenflo-asgari2021}. $S_8\equiv 
\sigma_8\,(\Omega_M/0.3)^{1/2}$, where $\sigma_8$ is the matter
clustering amplitude on the scale $8h^{-1}$\, Mpc (where $h$ is the Hubble
constant in units of 100\,km\,s$^{-1}$\,Mpc$^{-1}$). 

The main CMB observable that has been used to constrain the value of the
Hubble constant $H_0$ is the angular scale of the acoustic
peaks in the observed CMB spectrum. This scale can be characterized in
terms of an angle parameter $\theta_\star$ that has been determined from observations with an accuracy of about 0.03\,\%\ \cite{stenflo-planck2020a}. It is one of the best known parameters in cosmology. 

$\theta_\star$ is an angular measure of the anisotropy of the CMB radiation field, in which the observer is immersed. It is a local property of the radiation field in a comoving region. The angular scale is governed by basic physics that is directly related to the radius $r_\star$ of the sound horizon at the epoch $t_\star$ of hydrogen recombination. 

Parameters $\theta_\star$ and $r_\star$ do not contain information on the present value $H_0$ of the Hubble constant. The $H_0$ dependence enters when $\theta_\star$ and $r_\star$ are brought in relation with the so-called angular diameter distance $D_\star$ between the observer and the surface of last scattering via the defining equation 
\begin{equation}\label{eq:thetastar}
\theta_\star\equiv \frac{r_\star}{D_\star}\,,
\end{equation}
where 
\begin{equation}\label{eq:dstar}
D_\star = c\!\int_{t_\star}^{t_0}\frac{{\rm d}t}{a(t)}\,.
\end{equation}

So far there is no difference between our treatment and that of standard cosmology. The difference enters in the explicit expression for the $a(t)$ function, which determines how $D_\star$ is calculated. 

The anisotropy of the ambient radiation field in a comoving region depends on the evolution and dynamics of the local region. The $a(t)$ function in Equation (\ref{eq:dstar}) is therefore the scale factor in its dependence on the proper time that can be measured by a comoving clock. It does not contain any contribution from a $\Lambda$ term, because such a term only enters as an integration constant when we enforce the light-cone restriction to transform the unbounded 4D problem to a problem for a bounded and static 3D subspace. When the dynamics and evolution of a comoving region are calculated, the light-cone restriction is not applied or relevant. The $H_0$ tension arises when $\Lambda$ is not treated as an integration constant that is only relevant for the static 3D subspace, but is instead treated as a new field (dark energy) that also contributes to the scale factor evolution $a(t)$ of a comoving region. 

The explicit evaluation of the $D_\star$ integral is done in the same way as in Section \ref{sec:age} for the calculation of age $t_0$. The starting point is Equation (\ref{eq:atsolut}), which defines the solution for the $a(t)$ function that is used in Equation (\ref{eq:dstar}). The explicit expression for $D_\star$ then becomes 
\begin{equation}\label{eq:dstarexpl}
D_\star = \frac{c}{H_0} \int_{a_\star}^1 \frac{{\rm d}a}{a^2 \,(\Omega_M/a^3 +\Omega_R/a^4 ) ^{1/2}}\,,
\end{equation}
which is similar to that of $t_0$ in Equation (\ref{eq:age2}) except for the extra factor $a$ in the denominator, the factor $c$, and the lower integration limit $a_\star$ (at the epoch of last scattering) instead of zero (Big Bang). Note also (like in the $t_0$ case) that the appearance of the Hubble constant  in this expression is not because $H_0$ has any conceptual relation to $D_\star$, but because it comes from Equation (\ref{eq:omrho}), where it is used to define the conversion between the $\rho$ and $\Omega$ parameters. 

If we disregard $\Omega_R$, because it is $\ll \Omega_M$, then we get a simple approximate algebraic expression for $D_\star$ (similar to Equation (\ref{eq:age2})): 
\begin{equation}\label{eq:dstarapprox}
D_\star \approx \frac{c}{H_0\sqrt{1-\Omega_\Lambda}} \int_{a_\star}^1 \frac{{\rm d}a}{\sqrt{a}} \, \approx\frac{2c}{H_0\sqrt{1-\Omega_\Lambda}}\,.
\end{equation}
Our numerical evaluation, however, does not make use of this approximation but retains the $\Omega_R$ contribution and uses the full expression of Equation (\ref{eq:dstarexpl}). 

It is convenient to introduce the dimensionless parameter $x_\star$, which represents $D_\star$ in units of the Hubble radius $c/H_0$: 
\begin{equation}\label{eq:xstardef}
D_\star \,\equiv\, x_\star\,\frac{c}{H_0}\,.
\end{equation}
Converting for mathematical convenience the $a$ scale into a $z$ scale via Equation (\ref{eq:azdef}), we get 
\begin{equation}\label{eq:xstarexpr}
x_\star \,= \int_0^{z_\star} \frac{{\rm d}z}{ \,[\,\Omega_M(1+z)^3 +\Omega_R(1+z)^4\, ] ^{1/2}}\,.
\end{equation}
Note that this is a purely formal conversion, because no redshift observations are involved in the CMB analysis, and $z_\star$ is not an observable. $z$ is here a mathematical parameter that does not represent redshift. 

With Equations (\ref{eq:thetastar}), (\ref{eq:xstardef}), and (\ref{eq:xstarexpr}) it is possible to derive the value of $H_0$ from the observed anisotropies $\theta_\star$ in the CMB radiation field and the radius $r_\star$ of the sound horizon through 
\begin{equation}\label{eq:h0starderiv}
H_0 = c\,\, x_\star\, \theta_\star/r_\star\,.
\end{equation}
This results in a value for $H_0$ from CMB data that agrees with the value that has been derived from the redshift - brightness relation of supernovae observations (as shown explicitly below). Thus, with the current formalism there is no $H_0$ tension.

Equation (\ref{eq:h0starderiv}) also makes it clear why there must be an anti-correlation between the tensions in $H_0$ and the radius of the sound horizon $r_\star$ (or 
$r_d$), as noticed in \cite{stenflo-arendse2020}. Since $\theta_\star$ is fixed by the observations, the  
model dependence of the $H_0\,r_\star$ product is contained in the dimensionless $x_\star$
factor, which only depends on $\Omega_\Lambda$ if we assume spatial
flatness. For a given value of $\Lambda$ the variations in the $H_0$ and $r_\star$
parameters are therefore anti-correlated. 

The reason why CMB analysis with standard cosmology gives a different answer that is at odds with the supernovae results is that a different expression $x_{\rm st.\,cosm.}$ for the dimensionless parameter $x_\star$ is used. It is based on the implicit assumption of standard cosmology that the cosmological constant $\Lambda$ is a physical field that governs the behavior of both the $a(t)$ and $a(r)$ functions, rather than being a boundary condition that only affects the $a(r)$ function that represents the bounded 3D subspace. 

The $\theta_\star$ and $r_\star$ parameters are governed by the local evolution of the sound waves and the anisotropy of the electromagnetic radiation field, not by a distance-related function that depends on the location of an observer. The physics is therefore determined by the $a(t)$ function as in Equation (\ref{eq:dstar}), not by the $a(r)$ function. For this reason $x_\star$, as given by  Equation (\ref{eq:xstarexpr}), does not contain any $\Omega_\Lambda$ contribution. In contrast standard cosmology uses 
\begin{equation}\label{eq:xscexpr}
x_{\rm st.\,cosm.} \,= \int_0^{z_\star} \frac{{\rm d}z}{ \,[\,\Omega_M(1+z)^3 +\Omega_R(1+z)^4 + \Omega_\Lambda\, ] ^{1/2}}\,.
\end{equation}

When this expression is inserted in Equation (\ref{eq:h0starderiv}) instead of the expression (\ref{eq:xstarexpr}) for $x_\star$, one obtains a different value for the Hubble constant: 
\begin{equation}\label{eq:h0stcosmderiv}
H_{0,\,{\rm st.\,cosm.}} = c\,\, x_{\rm st.\,cosm.}\, \theta_\star/r_\star\,.
\end{equation}

The interpretation of the redshift - brightness relation for supernovae (SN) as standard candles must, on the other hand, be based on the $a(r)$ function for scale factor vs. distance along light rays rather than on the $a(t)$ function for comoving regions. It will therefore be affected by a boundary-induced integration constant that appears in the form of a $\Lambda$ term. As the existence of such a non-zero term is demanded by the observed SN relation between redshift and apparent brightness and is in fact being used for the interpretation of the observed  relation, it is the supernovae value $H_{0,\,{\rm SN}}$ that represents the correctly determined present Hubble constant. This allows us to make the identification 
\begin{equation}\label{eq:h0h0sn}
H_{0,\,{\rm  SN}}  \equiv H_0 \,,
\end{equation}
where $H_0$ is the correct value that appears in expression (\ref{eq:h0starderiv}). 

Because $\Omega_R$ is known from the observed temperature of the CMB radiation field, and $\Omega_M$ and $\Omega_\Lambda$ are related via Equation (\ref{eq:omnorm}), both $x_\star$ and $x_{\rm st.\,cosm.}$ are effectively functions of only $\Omega_\Lambda$ (if we disregard the minor model dependence on $z_\star$, the integration limit that represents the surface of last scattering). Dividing Equations (\ref{eq:h0starderiv}) and (\ref{eq:h0stcosmderiv}) with each other and using the identification of Equation (\ref{eq:h0h0sn}), we obtain an expression for the $H_0$ tension: 
\begin{equation}\label{eq:h0ratio}
\frac{H_{0,\,{\rm SN}}}{H_{0,\,{\rm st.\,cosm.}}}=\frac{x_\star(\Omega_\Lambda)}{x_{\rm st.\,cosm.} (\Omega_\Lambda)} \,. 
\end{equation}

With the theoretical value $\Omega_\Lambda\approx 0.671$, which was obtained through numerical solution of expression (\ref{eq:defomlam2}) for $\Omega_\Lambda$, evaluation of the full integral expressions (\ref{eq:xstarexpr}) and (\ref{eq:xscexpr}) for $x_\star$ and $x_{\rm st.\,cosm.}$ gives  
\begin{equation}\label{eq:h0ratiotheo}
\left(\frac{H_{0,\,{\rm SN}}}{H_{0,\,{\rm st.\,cosm.}}}\right)_{\!\!\rm theory}\!\approx
1.099 \,.
\end{equation}

This is to be compared with the observed value for the tension. According to the supernovae observations, $H_{0,\,{\rm SN}} =73.2\pm
1.3$\,km\,s$^{-1}$\,Mpc$^{-1}$ \cite{stenflo-riessetal2018,stenflo-riess2021a}  
, while CMB analysis in the framework of standard cosmology gives $H_{0,\,{\rm st.\,cosm.}} =67.4\pm
0.5$\,km\,s$^{-1}$\,Mpc$^{-1}$ \cite{stenflo-planck2020a}. 
The ratio between these two values represents the observed $H_0$ tension: 
\begin{equation}\label{eq:h0ratioobs}
\left(\frac{H_{0,\,{\rm SN}}}{H_{0,\,{\rm st.\,cosm.}}}\right)_{\!\!\rm obs}\!\approx
1.086\pm 0.021 \,.
\end{equation}
The theoretical prediction is thus well within $1\sigma$ of the observed value.

\section{Conclusions}\label{sec:concl}

Standard cosmology assumes that the cosmological constant represents some unknown form of physical field, commonly referred to as ``dark energy''. To explain the redshift - brightness observations of standard candles such a hypothetical field needs to have an equation of state that is nearly the same as the vacuum energy of quantum fields and a value, which in energy density units ``happens'' to be of the same order as the current matter energy density. This would suggest that our epoch in cosmic history is very special, a ``cosmic coincidence''. Another problem is that standard cosmology is inconsistent with respect to the Hubble constant $H_0$. The derived value for $H_0$ depends on the type of data that are being used, standard candles in the form of supernovae for the nearby universe, or CMB data for the early universe. 

Here we show that these stubborn enigmas and inconsistencies go away if one abandons the assumption that $\Lambda$ represents a physical field, and instead treats it as a covariant integration constant. This constant comes from a boundary condition that applies to the $a(r)$ function, the dependence of the scale factor $a$ on distance $r$ along light rays. While the $r$ scale is constrained to the bounded 3D hypersurface of the past light cone, which represents the observable universe, the proper time scale $t$  that can be measured in 4D spacetime by comoving clocks is unrelated to light cones. Consequently the $a(t)$ function cannot contain any boundary-induced integration constant. 

The integration constant has the dimension of the inverse square of a length scale. The only available characteristic length scale of the system is the limiting size $r_c$ of the observable universe. It is therefore not surprising that the value of the cosmological constant is linked to the $r_c$ scale that characterizes the universe now. This means that our epoch is not special, there is no violation of the Copernican principle 

When the distinction between the $a(r)$ and $a(t)$ functions is accounted for, the $H_0$ tension goes away. While the Hubble constant that has previously been determined from the redshift - brightness observations of supernovae as standard candles remains unaffected, our revised CMB analysis gives a value for $H_0$ that now agrees with the supernovae value, in contrast to the CMB value that is obtained within the interpretational framework of standard cosmology. 

Another consequence is that the proper age of the universe gets increased by 1.7\,Gy, from the 13.8\,Gyr of standard cosmology to 15.5\,Gyr. This is consistent with other age determinations and in fact relieves some tension with ages that have been determined for the oldest halo stars in the Milky Way. The error bars of stellar ages however need to come down before they can be used reliably for the discrimination between the cosmological frameworks. 

So is the universe accelerating or not according to the observational data\,? Our answer is no, it is decelerating, because the $a(t)$ function does not contain any cosmological constant. It is the $\Lambda$-free proper time scale $t$ that is relevant for the dynamics. Standard cosmology answers yes, because it interprets the $a(r)$ function that contains a non-zero $\Lambda$ term in terms of dynamics. However, while distances $r$ can formally be expressed in time units as  ``look-back time'', this time scale is not relevant for dynamics and therefore not for the question of an accelerated expansion.



\vspace{6pt} 

\funding{This research received no external funding.}

\dataavailability{Not applicable.}

\acknowledgments{I am grateful to \AA ke Nordlund for many helpful suggestions.}

\conflictsofinterest{The author declares no conflict of interest.}

\begin{adjustwidth}{-\extralength}{0cm}

\reftitle{References}


\end{adjustwidth}
\end{document}